\documentclass[twocolumn,aps,prd]{revtex4-1}
\usepackage{graphicx}
\usepackage{times}
\usepackage{color}
\usepackage{float}
\usepackage{amsmath}
\graphicspath{{figures_witness/}{./}}
\usepackage[T1]{fontenc}
\usepackage[utf8]{inputenc}
\usepackage{amssymb}
\usepackage{amsfonts}
\usepackage{amsmath}
\usepackage{amsbsy}
\usepackage{natbib}
\usepackage{comment}
\usepackage{amssymb}
\usepackage{color}
\usepackage{float}
\usepackage{comment}
\usepackage{mdframed}

\newcommand{\homega}{\bar \omega}

\newcommand{\mr}{\mathrm}

\newcommand{\bra}[1]{\left\langle #1 \right\vert}
\newcommand{\ket}[1]{\left\vert #1 \right\rangle}

\begin{document}


\title{Quantum witness and invasiveness of cosmic neutrino measurements}

%
\author{P. Kurashvili$^{1}$, L. Chotorlishvili$^{2}$, K. A. Kouzakov$^{3}$, A. G. Tevzadze$^{4}$, A. I. Studenikin$^{5,6}$}
\address{$^{1}$ National Centre for Nuclear Research, Warsaw 00-681, Poland\\
$^{2}$ Institute f\"ur Physik, Martin-Luther Universit\"at Halle-Wittenberg, D-06120 Halle/Saale, Germany\\
$^{3}$ Department of Nuclear Physics and Quantum Theory of Collisions, Faculty
of Physics, Lomonosov Moscow State University, Moscow 119991, Russia\\
$^{4}$ Kutaisi International University, Kutaisi University Campus, Kutaisi 4600, Georgia\\
$^{5}$ Department of Theoretical Physics, Faculty of Physics,
Lomonosov Moscow State University, Moscow 119991, Russia\\
$^{6}$ Joint Institute for Nuclear Research, Dubna 141980, Moscow Region,
Russia}
\date{\today}
%

%
\begin{abstract}
Measurements of cosmic neutrinos have a reach potential for providing an insight into fundamental neutrino properties. For this a precise knowledge about an astrophysical environment of cosmic neutrinos propagation is needed. However this is not always possible, and the lack of information can bring about theoretical uncertainties in our physical interpretation of the results of experiments on cosmic neutrino fluxes. We formulate an approach that allows one to quantify the uncertainties using the apparatus of quantum measurement theory. We consider high-energy Dirac neutrinos emitted by some distant source and propagating towards the earth in the interstellar space. We take into account the neutrino magnetic moment interaction with stochastic interstellar magnetic fields and describe the neutrino evolution in the formalism of the Lindblad master equation for the neutrino density matrix. It is supposed that neutrinos can meet on their way to the detector at the earth a dense cosmic object serving as a filter that ``stops'' active, left-handed neutrinos and letting only sterile, right-handed neutrinos to propagate further. Such a filter mimics the strongest effect on the neutrino flux that can be induced by the cosmic object and that can be missed in the theoretical interpretation of the lab measurements due to the insufficient information about the astrophysical environment of the neutrino propagation. Treating the neutrino interaction with the cosmic object as the first, neutrino-spin measurement, whose result is not recorded, we study its invasive effect on the second, neutrino-flavor measurement in the lab. We quantify the invasiveness of the first, blind measurement by means of quantum witness that in the discussed case has an advantage over the well-known Leggett–Garg inequality, since the latter explores two-time correlation functions of the same operator. We solve analytically the Lindblad master equation for time evolution of the neutrino density matrix and on this basis we calculate the quantum witness for measuring cosmic electron neutrinos in the lab. We present numerical illustrations of the invasive effect showing that the quantum witness as a function of the distance between the earth and cosmic object can be asymptotically nonvanishing despite the neutrino spin thermalization induced by stochastic interstellar magnetic fields.
\end{abstract}
\maketitle
\section{Cosmic neutrinos and the concept of quantum measurement}
Many objects in the Universe produce vast amount of cosmic neutrinos with
different energies. These include highly relativistic particles from high
energy (>100~MeV) to extremely high energy (>1PeV) or even ultra  high energy
(>1~EeV) range. Neutrinos carry unique information about the internal
environment of high energy astrophysical objects that are not normally
directly observable.

It is known that 99\% of the energy of a core collapse supernova explosion
is carried away by a neutrino flux. All these particles contribute to the
diffuse supernova neutrino background observed on the earth~\cite{Zhang2015}.
Still the first albeit only detection of neutrino burst from a single
supernova explosion was detected for the 1987 event that occurred 51~kpc
away~\cite{Hirata1987}. The neutrino (antineutrino) flux from this extragalactic
explosion has arrived before the visible light of the explosion has reached
the earth. Neutrinos from the core of the proto-neutron star have travelled
away from strongly magnetized interior and were emitted towards the earth.

Magnetic moments of the neutrinos emitted from the core-collapse
supernova should be aligned towards the magnetic field axis of the
developing neutron star. Strongest polarization should occur for neutrinos
emitted from magnetars, where a magnetic field can reach up to
$10^{14}$-$10^{15}$~G. These objects are developed from at least $10\%$ of
the core collapse supernova and are known for strongest magnetic fields in the
Universe~\cite{Kaspi2017}.

Even stronger flux of magnetically polarized neutrinos should originate from
the neutron star mergers. Two cases of this phenomenon, long thought to be an
exotic event, is already detected using the gravitational wave
signal~\cite{Abbott2020}. Simultaneous detection of the gravitational wave and
neutrino flux will allow us to employ these particles in multi-messenger astronomy. The list of exotic events producing high-energy neutrino fluxes in strong magnetic fields may include hypothesized quark-nova and
collapsars.

At the extra high-energy scales neutrinos are born during Gamma Ray Bursts
from hypernova \cite{Wang2007}. Similar energies are expected to be emitted by
quasars in the active galactic nuclei, where supermassive black holes create
relativistic outflows~\cite{Kalashev2015,Aab2020}. With much higher
uncertainties of local properties of these objects it is clear that a strong
magnetic field should be present near the active area~\cite{Kollatschny2015}.
Thus, these luminous sources of neutrinos can be detected for extragalactic
distances exceeding 1 Gpc.

Neutrino fluxes reaching us from such distant sources are influenced by galactic
and extragalactic magnetic fields. These weak fields can exert a dissipative
effect on the neutrino spin polarization by their stochastic
components. Although microscopic in amplitudes these fields affect particles
at cosmic distances and can have a significant cumulative effect. Magnetic
fields in our Galaxy do not exceed 4~$\mu$G on average~\cite{Sobey2019},
though are stronger at the galactic center.

Magnetic fields are detected at extragalactic scales as well. These fields can
reach $10^{-9}$~G amplitudes at 1~Mpc scale~\cite{Ade2016}. Assuming
a turbulent spectrum of extragalactic magnetic fields we may estimate small
scale stochastic component to reach up to 0.3~$\mu$G
amplitudes~\cite{Brandenburg2017}. These weak fields can affect ultra high-energy neutrinos that travel up to Gpc distances.

From the plethora of cosmic neutrino emitting objects we focus on such sources
that can be found in the high-energy as well as in extra high-energy intervals.
Below, we consider evolution of neutrinos that are emitted from a distant source and traverse the cosmic space  diluted with the stochastic magnetic fields until they reach the detector in the lab.
While propagating in a magnetic field, neutrinos experience spin and spin-flavor oscillations due to a nonzero neutrino magnetic moment~\cite{bib:Giunti,andp16}. For simplicity we limit ourselves with the case of two neutrino flavors ($\nu_{e}$ and $\nu_{\mu}$).

In this paper we propose a new tool for the theoretical analysis of cosmic neutrino measurements, which
is based on the theoretical framework for quantum witness experiments.
For this we define two consequent quantum measurements of a neutrino.
Suppose that neutrinos meet a high-density cosmic object
on their way from the  source to the detector. 
Then, only part of neutrinos in the flux crosses the cosmic object without loss:
mainly right-handed neutrinos would pass the object, while left-handed neutrinos would be deflected or absorbed. Setting this effect
as the quantum spin measurement, then we can use the neutrino flavor detection at the lab
as the second measurement.


We describe the measurements by the positive-operator valued measure projectors. The first projector measurement is the essence of the quantum Maxwell Demon \cite{Deffner}, which holds the information, whether neutrinos of particular types get in and out of the cosmic filter. The second projector measurement is a typical neutrino experiment in the lab that defines the neutrino flavor. However, the result of the second measurement incisively depends on the outcome of the first measurement: our interpretation of the result (for example, such as the origin of neutrinos, the initial neutrino flux, its flavor and spectral composition, etc.) can be critically affected by whether Demon shares his results with us or not. In the latter case, the first measurement appears to be a blind measurement, i.e., its results are not recorded. Clearly, this mimics the situation when we do not have enough information about the conditions of cosmic neutrino propagation, in particular, about the presence of the cosmic object. Since the lack of information can lead to an incorrect physical interpretation of cosmic neutrino measurements in the lab, the approach developed in this work allows us to quantify the credibility of our interpretation. 

In what follows, we analyze cosmic neutrinos in the interstellar space filled with a magnetic field and explore the invasive effect of the cosmic object on the neutrino measurement in the lab. The standard tool for analyzing the invasiveness of quantum measurements is the Leggett–Garg method.  For neutrino physics, Leggett–Garg inequality was studied only recently in the case of unitary evolution \cite{Formaggio}. However, in the case of open quantum systems, exploring the two-time correlation functions is a formidable problem \cite{Shibata}, analytically not accessible for the neutrino problem. Therefore, as an alternative to the Legget-Garg method, we exploit the concept of quantum witness and the Novikov's dissipative channel \cite{Novikov}.

It should be noted that the phenomenon of neutrino oscillations is closely related to the concept of quantum coherence.  This concept touches upon the macroscopic coherence, i.e., superposition of a macroscopic large number of states and involves two core principles: (i)  macroscopic realism per se argues that the pre-existing value of the quantity in question can be inferred through the measurements done on the macroscopic system, (ii)
noninvasive measurability means that one can perform measurement without distorting the state of the system \cite{EmaryLambertNori, HoffmannSpee, KoflerBrukner, HuangJiajieLing,
HalliwellMawby,BanerjeeJayannavar,NikitinToms,NathanWilliamsAndrew,MeschedeEmary,SkinnerRuhmanNahum}.
In reality the situation is more complex.
Except for specific initial states, the measurement has a back-action and induces noncommuting dynamical changes in an observable
\cite{Leggett}. When the system initially is prepared in the superposition of two (or more) states $|\phi\rangle=a|\psi_{n}\rangle+b|\psi_{m}\rangle$, using measurement operator $\hat{\prod}_{n}=|\psi_{n}\rangle\langle\psi_{n}|$, $n\in \mathcal{N}$, one cannot determine the state of the system without a destructive effect on the state (the invasive measurement).  The quantum witness quantifies the invasiveness of the measurement. Up to date in quantum metrology mainly  nonrelativistic quantum systems have been discussed. Nevertheless, the interest to studies of such relativistic systems  as neutrinos   has recently emerged \cite{Formaggio,HuangJiajieLing}.


Below we elaborate on the general formalism for describing the quantum witness of the cosmic neutrino measurement  and discuss its application in feasible neutrino experiments.
We treat the neutrino evolution using the method based on the Lindblad master equation \cite{Lindblad}.
This approach is presently widely used in studies of neutrino quantum decoherence in different environments and under various  experimental conditions
(see, for instance, Refs. \cite{Stankevich:2020icp,Farzan_Schwetz_Smirnov,Lisi_Marrone,Barenboim_Mavromato,Barenboim_Mavromatos2,	 Benatti_Floreanini,Oliveira2014,Oliveira2016,
Balieiro_Guzzo,Joao_Coelho,Joao_Coelho2,Capolupo_Giampalo,Coloma}.

The work is organized as follows. In Sec.~\ref{gen_form}, the general formulation is presented. In Sec.~\ref{sec:quantum_witness} we define the quantum witness for the neutrino flavor measurement in the lab. Section~\ref{subsec:ProjectionOperators} is devoted to the analytical derivation of the quantum witness from the neutrino density matrix. In Sec.~\ref{sec:AnalysisAndConclusions}, we give illustrations of the invasive effect of the cosmic object on the neutrino flavor measurement in the lab. Section~\ref{summary} summarizes this work. In Appendix, we deliver details of solving the Lindblad master equation for neutrino evolution.

%
\section{Neutrino evolution in an interstellar magnetic field}
\label{gen_form}
In the scope of our interest are two Dirac neutrino helicity basis states
$\vert\nu_{1}^{\pm }\rangle$, $\vert\nu_{2}^{\pm }\rangle$ with masses
$m_1$ and $m_2$. For the sake of convenience
we switch from the mass basis $(\nu_{1}^{+},\nu_{1}^{-},\nu_{2}^{+},\nu_{2}^{-})^T$ to the flavor basis $(\nu_{e}^{R},\nu_{e}^{L},\nu_{\mu}^{R},\nu_{\mu}^{L})^T$:
\begin{eqnarray}\label{flavor basis}
\vert\nu_{e}^{R,L}\rangle&=&\cos\theta\vert\nu_{1}^{\pm }\rangle+\sin\theta\vert\nu_{2}^{\pm } \rangle,\nonumber\\
\vert\nu_{\mu}^{R,L}\rangle&=&-\sin\theta \vert\nu_{1}^{\pm }\rangle+\cos\theta\vert\nu_{2}^{\pm } \rangle.
\end{eqnarray}
The Hamiltonian of the problem is given by (see Ref. \cite{Kurashvili})
\begin{eqnarray}\label{Hamiltonian of the problem}
\hat{H}_{eff}=\hat{H}_{vac}+\hat{H}_{B},
\end{eqnarray}
where $\hat{H}_{vac}$ is the vacuum part %
\begin{equation}
\label{eq:HamiltonianFlavorRepresentation}
{\hat H}_{vac}=\omega_\nu
\begin{pmatrix}
-\cos 2 \theta & 0 & \sin 2 \theta & 0
\\
0 & -\cos 2 \theta & 0 & \sin 2 \theta
\\
\sin 2 \theta & 0 & \cos 2\theta & 0
\\
0 & \sin 2 \theta & 0 & \cos 2\theta
\end{pmatrix},
\end{equation}
with
\begin{equation}
\label{eq:DeltaM}
\omega_\nu= \frac{\Delta m^2}{4 E_\nu}, \qquad \Delta m^2 =m_2^2 - m_1^2,
\end{equation}
and $E_\nu$ being the neutrino energy. 
The Hamiltonian of the neutrino interaction with a magnetic field in the
flavor representation can be presented as~\cite{bib:Fabbricatore}
\\
\begin{widetext}
\begin{equation}
\label{eq:H_EM}
H_{B}=
\begin{pmatrix}
\displaystyle -\left(\frac{\mu}{\gamma}\right)_{ee} {B_\parallel} &&
\mu_{ee}B_{\perp} &&
\displaystyle -\left(\frac{\mu}{\gamma}\right)_{ e\mu}{B_\parallel} &&
\mu_{e\mu}B_\perp
\\
\mu_{ee}B_\perp &&
\displaystyle -\left(\frac{\mu}{\gamma}\right)_{ee}{B_\parallel} &&
\mu_{e\mu} B_\perp &&
\displaystyle -\left(\frac{\mu}{\gamma}\right)_{e\mu}{B_\parallel}
\\
\displaystyle -\left(\frac{\mu}{\gamma}\right)_{e\mu}{B_\parallel} &&
\mu_{e\mu}B_{\perp} &&
\displaystyle -\left(\frac{\mu}{\gamma}\right)_{\mu \mu}{B_\parallel}
&&
\mu_{\mu \mu} B_\perp
\\
\mu_{e\mu}B_{\perp} &&
\displaystyle -\left(\frac{\mu}{\gamma}\right)_{e\mu}{B_\parallel} &&
\mu_{\mu \mu}B_\perp &&
\displaystyle -\left(\frac{\mu}{\gamma}\right)_{\mu\mu}{B_\parallel}
\end{pmatrix},
\end{equation}
\end{widetext}
where $B_\parallel$ and $B_\perp$ are the parallel and transverse magnetic-field components with respect to the neutrino velocity, and the magnetic moments $\tilde{\mu}_{\ell\ell'}$ and $\mu_{\ell\ell'}$ ($\ell,\ell'=e,\mu$) are related to those in the mass representation $\mu_{jk}$ ($j,k=1,2$) as follows:
\begin{align}
\label{eq:MuPrime}
\mu_{ee}&=\mu_{11} \cos^2 \theta +\mu_{22} \sin^2 \theta
+\mu_{12} \sin 2\theta,
\nonumber
\\
\mu_{e\mu}&=\mu_{12}\cos 2\theta + \frac{1}{2}
\left( \mu_{22} - \mu_{11}\right)\sin 2\theta,
\\
\mu_{\mu\mu}&=\mu_{11} \sin^2 \theta
+\mu_{22} \cos^2 \theta-\mu_{12} \sin 2\theta,
\nonumber
\end{align}
and
\begin{align}
\label{eq:MuTilde}
\left(\frac{\mu}{\gamma}\right)_{ee} &=
\frac{\mu_{11}}{\gamma_{1}}\,\cos^2 \theta +
\frac{\mu_{22}}{\gamma_{2}}\,\sin^2 \theta +
\frac{\mu_{12}}{\gamma_{12}}\,\sin 2\theta,
\nonumber
\\
\left(\frac{\mu}{\gamma}\right)_{e\mu} &=
\frac{\mu_{12}}{\gamma_{12}}\,\cos 2\theta
+\frac{1}{2}\left(
\frac{\mu_{22}}{\gamma_{2}}-\frac{\mu_{11}}{\gamma_{1}}
\right)\sin 2\theta,
\\
\left(\frac{\mu}{\gamma}\right)_{\mu\mu}&=
\frac{\mu_{11}}{\gamma_{1}}\,\sin^2 \theta
+\frac{\mu_{22}}{\gamma_{2}}\,\cos^2 \theta
-\frac{\mu_{12}}{\gamma_{12}}\,\sin 2\theta.
\nonumber
\end{align}
Here $\gamma_1$ and $\gamma_2$ are the Lorenz factors of the massive neutrinos, and
\begin{equation}
\label{eq:GammaDefinition}
\frac{1}{\gamma_{12}}=\frac{1}{2}\left(\frac{1}{\gamma_1}+\frac{1}{\gamma_2}\right).
\end{equation}
We consider the case when the galactic and extragalactic magnetic fields are composed of
the large-scale regular component $\vec{B}$ that enters Eq. (\ref {eq:H_EM})
and a small-scale stochastic component $\vec{h}$.

The stochastic  magnetic field $\vec{h}$ is a result of interstellar
fluctuations, galactic winds, cosmic turbulence and primordial magnetic field
fluctuations.
It is characterized by the correlation function~\cite{Garanin:1997prb}
$\langle h_\alpha(t)h_\beta(0)\rangle=\frac{w^2}{2\mu_\nu^2}\delta(t)$,
where $\mu_\nu$ is a putative neutrino magnetic moment and $w^2=k_BT$, with $T$ being the effective temperature.


The density matrix of the system obeys the Lindblad master equation \cite{Lindblad} in the form:
\begin{eqnarray}\label{master equation}
 \frac{d\hat{\varrho}}{dt}=-{i}\left[\hat{H},\hat{\varrho}\right]
-\frac{w^{2}}{2}\left(\hat{\varrho}\hat{{V}}^2+\hat{V}^2\hat{\varrho}-2\hat{{V}}\hat{\varrho}\hat{{V}}\right).
\end{eqnarray}
In the most general case the $\hat{{V}}$ matrix in Eq.(\ref{master equation}) is given by
\begin{equation}
      \label{Vmatrix}
      V_{ik} = \bra i \hat{I}^{(\nu_e)}  \otimes \hat{v}^{(\nu_\mu)} + \hat{I}^{(\nu_\mu)} \otimes \hat{v}^{(\nu_e)} \ket k ,
\end{equation}
where $\ket i$ and $\ket k$ ($i,k = 1,2,3,4$) are the eigenstates of the Hamiltonian $\hat{H}_{eff}$ (see Ref. \cite{Kurashvili} for details). $\hat{I}^{(\nu_\ell)}$ is the $2\times 2$ unit matrix acting in the Hilbert space of the $\nu_\ell$ neutrino. The  $2\times 2$ matrix $\hat{v}^{(\nu_\ell)}$ also acts in the Hilbert space of the $\nu_\ell$ neutrino. It can be presented as
\begin{equation}
\label{eq:vsigma_0}
\hat{v}^{(\nu_\ell)} = v_0 \hat{I} + \vec v \cdot \hat{\vec\sigma},
\end{equation}
where $\hat{\vec\sigma}$ is the Pauli vector. Below we utilize the following parametrization:
\begin{equation}
\vec{v} = (v \sin \beta \cos \alpha,
v \sin \beta \sin \alpha, v \cos \beta).
\label{parametrization}
\end{equation}
In its general form the dissipator~(\ref{Vmatrix}) describes relaxation of both transverse
and longitudinal neutrino spin components. However, in the particular case
$\beta=\pi/2$ the cosmic magnetic field does not thermalize the $\sigma_z$ component
of the neutrino spin.

We analytically solve Eq.~(\ref{master equation}) in the
eigenbasis of the Hamiltonian  (\ref{Hamiltonian of the problem}). The solution $\hat{\varrho}(t)$
is cumbersome and is presented in Appendix~\ref{sec:app_lindblad}.

\section{Quantum witness of cosmic neutrino measurements}
\label{sec:quantum_witness}
As was already mentioned, if neutrinos pass through the cosmic
object they acquire a preferential helicity polarization, since the high-density matter filters out the left-handed neutrinos. This process can be described by the positive-operator valued measure (POVM)
projectors, projecting the neutrino state on the direction of the neutrino flux propagation.

Let us first consider the case when neutrinos pass through a dense cosmic object (we call this
``the first propagation scheme''). The initial neutrino state at $t=0$, i.e., just before entering the object, is  $|\phi\rangle$. 
In the general case the efficient quantum
measurement of neutrino spin polarization transforms this state into the  post measurement state
\begin{eqnarray}
\label{post-measurement01}
|\Phi \big\rangle=\frac{\big(\hat{\Pi}_{\vec s}\bigotimes \hat{I}^{(\ell)}\big)
\big|\phi \big\rangle}{\sqrt{\big\langle \phi \big|\big(\hat{\Pi}_{\vec s}\bigotimes
\hat{I}^{(\ell)}\big)\big|\phi \big\rangle}},
\end{eqnarray}
where $\vec s$ is a unit vector of spin polarization of the neutrino on the way out of the cosmic object 
and $\hat{I}^{(\ell)}$ is the identity operator acting on the flavor space.
Taking into account that the left-handed neutrinos are mostly filtered-out by the cosmic object, the post measurement density matrix is given by
\begin{eqnarray}\label{2postdensity matrix}
\hat{\varrho}_{\mr{post}}=|\Phi\big\rangle\big\langle\Phi |= 
\frac{(\hat{\Pi}_{+}\bigotimes \hat{I}^{(\ell)})\hat{\varrho}(\hat{I}^{(\ell)}\bigotimes \hat{\Pi}_{+})}{\mr{Tr} \left((\hat{\Pi}_{+}\bigotimes \hat{I}^{(\ell)})
\hat{\varrho}(\hat{I}^{(\ell)}\bigotimes\hat{\Pi}_{+})\right)},
\end{eqnarray}
where $\hat{\varrho}=|\phi \rangle\langle\phi \vert$ is the initial
density matrix and the positive-helicity projector operator is
\begin{equation}
\hat{\Pi}_{+} = \frac{1}{2}\left(1 + \frac{\vec p \vec \sigma }{|\vec p|}\right).
\label{eq:p_projection_spin_plus}
\end{equation}


After measuring helicity we evolve the density matrix through the
trace-preserving Novikov's map $\mathcal{\hat{F}}\left[\hat{\varrho}_{\mr{post}}\right] $
that mimics the effect of a stochastic magnetic field in Eq. (\ref{master equation}).
The second measurement is then performed by detecting the active flavor neutrino state $\vert\nu_\ell^L\rangle$ in the lab.
We describe this detection procedure through the projector operator
$\hat{\Pi}_{\nu_\ell^L}= \hat{\Pi}_{-}\hat{\Pi}_{\ell}$, where the
negative-helicity projector operator reads
\begin{equation}
\hat{\Pi}_{-} = \frac{1}{2}\left(1 - \frac{\vec p \vec \sigma }{|\vec p|}\right),
\label{eq:p_projection_spin_minus}
\end{equation}
and the flavor projector operator is $\hat{\Pi}_{\ell}=\vert \ell\rangle\langle \ell\vert$, $\ell =
e,\mu$. Thus, the probability of detecting the active flavor neutrino state $\vert\nu_{\ell}^L\rangle$  is given by
\begin{eqnarray}\label{direct spin measurement4}
\mathcal{Q}^{(\ell)}_{L}=\mr{Tr}\{\hat{\Pi}_{\nu_{\ell}^L}\mathcal{\hat{F}}[\hat{\varrho}_{\mr{post}}]\}.
\end{eqnarray}
We now consider the second propagation scheme, meaning that the neutrinos do not meet the cosmic object on its way from the source to the detector. In this case, the neutrino flavor state is measured without preliminary measurement of the neutrino helicity.
The probability of detecting the active flavor neutrino state $\vert\nu_{\ell}^L\rangle$ is given by
\begin{eqnarray}\label{direct spin measurement2}
\mathcal{P}^{(\ell)}_{L}=\mr{Tr}\{\hat{\Pi}_{\nu_{\ell}^L}\mathcal{\hat{F}}[\hat{\varrho}]\}.
\end{eqnarray}

The difference between the two neutrino propagation schemes is due to the invasive effect of the neutrino helicity measurement ``performed'' by the cosmic object.  For quantifying the invasiveness of the neutrino measurement in the lab we use the quantum witness
\begin{eqnarray}\label{quantum witness2}
\mathcal{W}_{L}^{(\ell)}=\big|\mathcal{P}^{(\ell)}_{L}-\mathcal{Q}^{(\ell)}_
{L}\big|.
\end{eqnarray}
Note that quantum witness ranges from 0 to 1, so that the value of 0 corresponds to no invasive effect and that of 1 to a maximal invasive effect of the first measurement performed by the cosmic object. Accordingly, the confidence in the interpretation of the result of the second measurement is either minimal or maximal depending on whether quantum witness~(\ref{quantum witness2}) equals 0 or 1.  


\section{Quantum witness and the density matrix}
\label{subsec:ProjectionOperators}
In the mass basis, the entire $4\times 4$ density matrix can be presented in the conventional form
\begin{equation}
\hat \varrho = 
\begin{pmatrix}
\varrho^{(1)} & \varrho^{(2)} \\ \varrho^{(3)} & \varrho^{(4)}
\end{pmatrix},
\label{eq:dmatrix_minors}
\end{equation}
where $\varrho^{(\alpha=1,2,3,4)}$ are the matrices of dimension $2\times 2$.
The elements of the four quadrants
can be enumerated with separate sets of indices
$1$ and $2$, corresponding to the spin-up and
spin-down states, respectively.
The quadrants we expand over the basis of Pauli matrices:
\begin{equation}
\label{eq:rsigma}
\varrho^{(\alpha)} = r^{(\alpha)}_0 I +  \vec \sigma \vec r^{(\alpha)}.
\end{equation}
The elements of matrices $\varrho^{(\alpha)}$
are linked to the respective coefficients
$r^{(\alpha)}_{i=0,1,2,3}$ through the following relations:
\begin{align}
\label{eq:varrho11}
\varrho^{(\alpha)}_{11} & = r^{(\alpha)}_0 + r^{(\alpha)}_3,
\\
\label{eq:varrho22}
\varrho^{(\alpha)}_{22} & = r^{(\alpha)}_0 - r^{(\alpha)}_3,
\\
\label{eq:varrho12}
\varrho^{(\alpha)}_{12} & = r^{(\alpha)}_1 - i r^{(\alpha)}_2,
\\
\label{eq:varrho21}
\varrho^{(\alpha)}_{21} & = r^{(\alpha)}_1 + i r^{(\alpha)}_2.
\end{align}
In these variables, the entire  Lindblad equation for the density matrix splits into a set of four independents linear systems. Each quadrant contains four elements and admits the exact analytical solution  $r^{(\alpha)}_i(t)$.
We evolve the density matrix through the trace-preserving dissipative channel, conserving the entire density matrix's trace. Therefore the sum of traces of the diagonal quadrants $\varrho^{(1)}$ and
$\varrho^{(4)}$ is the integral of motion.

The measurement is the essence of the action
of flavor and spin projection matrices on the density matrix. In terms of Eq. (\ref{eq:dmatrix_minors}), the effect of the application
of the spin projection operator is expressed as the action of the operator
$\hat{\Pi}_\pm$ on both sides of the $2\times 2$ matrix
minors $\varrho^{(\alpha)}$, where $\alpha$ is either
$1$ or $4$:
\begin{align}
\label{multiplicationprojection}
\hat \Pi_{\vec s} \varrho^{(\alpha)}(0) \hat {\Pi}_{\vec s}  = &
 \frac{1}{4} (1 +  \vec \sigma \vec{s})
\left[r^{(\alpha)}_0(0) I + \vec{\sigma}\vec{r}^{(\alpha)}(0)\right] \nonumber\\
&\times(1  + \vec \sigma \vec{s}) \nonumber\\
=& \frac{1+ \vec{\sigma}\vec{s}}{2}
\left[r^{(\alpha)}_0(0) + \vec{r}^{(\alpha)}(0)\vec{s}\right].
\end{align}
The respective traces read:
\begin{align}
\label{eq:Trace_plus}
N_+ & = \mathrm{Tr}(\hat \Pi_+ \varrho^{(1)}(0) \hat \Pi_+) +
\mathrm{Tr}(\hat \Pi_+ \varrho^{(4)}(0) \hat \Pi_+)
 \nonumber\\
&= r_0^{(1)}(0)+r_3^{(1)}(0) + r_0^{(4)}(0)+r_3^{(4)}(0)
 \nonumber\\
&=  \frac 1 2 + ( r^{(1)}_3(0)+ r^{(4)}_3(0))
\end{align}
and
\begin{align}
\label{eq:Trace_minus}
N_- & = \mathrm{Tr}(\hat \Pi_- \varrho(0) \hat \Pi_-) +
\mathrm{Tr}(\hat \Pi_- \varrho^{(4)}(0) \hat \Pi_-)
 \nonumber\\
& = r_0^{(1)}(0)-r_3^{(1)}(0) + r_0^{(4)}(0)-r_3^{(4)}(0)
 \nonumber\\
& = \frac 1 2 - ( r^{(1)}_3(0)+ r^{(4)}_3(0)).
\end{align}
Note that in terms of components of the full $4\times 4$ matrix,
$\hat \varrho$,
$N_+ = \hat \varrho_{11}(0) + \hat \varrho_{33}(0)$,
$N_- = \hat \varrho_{22}(0) + \hat \varrho_{44}(0)$.

The post measurement density matrix quadrants are equal to
\begin{align}
\label{eq:post_matrix_plus}
\varrho_{\mr{post}}^{(\alpha)}(0) & =
\frac 1 2  \frac{(r_0^{(\alpha)}(0) + r_3^{(\alpha)}(0))(1  +  \sigma_z )}
{N_+}
 \nonumber\\
& = \frac{ r_0^{(\alpha)}(0) + r_3^{(\alpha)}(0)}
{ 1 + 2 (r_3^{(1)}(0)+r_3^{(4)}(0))}
(1 + \sigma_z).
\end{align}
%
The post-measurement density matrices
$\varrho^{(\alpha)}_{\mr{post}}$ obey the equation of motion but
for different initial conditions as compared to the case 
when the neutrino density matrix is not filtered out through the first 
helicity measurement.  
The coefficient $r^{(\alpha)}_0$ of the expansion in Eq. (\ref{eq:rsigma})
is, in the essence, the trace of a $2\times2$ matrix $\varrho^{(\alpha)}$
and is conserved in time. 

Let us derive the projection operators 
for the electron neutrino in the explicit form:
\begin{align}
\hat \Pi_{\nu_e^{R, L}} = & 
\vert \nu_e^{R, L} \rangle \langle \nu_e^{R, L} \vert\nonumber
\\
 = &
\left( \cos \theta\vert \nu_1^\pm \rangle + 
 \sin \theta \vert \nu_2^\pm\rangle \right) \nonumber\\ 
&\times
 \left(\langle \nu_1^\pm \vert  \cos \theta + 
\langle \nu_2^\pm \vert \sin \theta \right).
\label{eq:projection_electron}
\end{align}
Following the same recipe, one can derive the POVM projectors for the muon neutrino.
Therefore we deduce for the flavour projectors in the mass basis:
%
\begin{align}
\hat \Pi_{e} = 
\begin{pmatrix}
c_\nu^2 && c_\nu s_\nu \\
c_\nu s_\nu && s^2_\nu
\end{pmatrix}
\label{eq:projection_e_flavor}
\end{align}
and
\begin{align}
\hat \Pi_{\mu} = 
\begin{pmatrix}
s_\nu^2 && - c_\nu s_\nu \\
- c_\nu s_\nu && c^2_\nu
\end{pmatrix},
\label{eq:projection_mu_flavor}
\end{align}
where we introduced the notations $c_\nu = \cos \theta$, $s_\nu = \sin \theta$.
In what follows, we exploit the flavour projection operators in the form:
\begin{align}
\label{eq:projection_e_flavor_alt}
\hat \Pi_{e}& = \frac 1 2 \left( 1 + \sigma_1 \sin 2 \theta
+ \sigma_3 \cos 2 \theta \right), 
\\
\label{eq:projection_mu_flavor_alt}
\hat \Pi_{\mu}& = \frac 1 2 \left( 1 - \sigma_1 \sin 2 \theta
- \sigma_3 \cos 2 \theta \right).
\end{align}

We insert these operators into the expression for the result in  
the second measurement scheme~(\ref{direct spin measurement2}):
\begin{align}
\label{eq:direct_left}
\mathcal{P}^{(\ell)}_{L} &= 
\mr{Tr}\{ \hat{\Pi}_{\nu_\ell^L}\hat{\mathcal{F}}[\hat \varrho] \} 
=\mr{Tr}\{\hat{\Pi}_{\nu_\ell^L}\hat \varrho(t)\hat{\Pi}_{\nu_\ell^L}\} 
\nonumber\\
& =\mathrm{Tr} \{\hat \Pi_- \hat \Pi_{\ell} \hat\varrho(t) 
\hat \Pi_{\ell} \hat \Pi_- \}.
\end{align}
Employing the density matrix in the form 
Eq.~(\ref{eq:dmatrix_minors}) and applying the spin and flavor projection 
operators given by Eqs.~(\ref{eq:p_projection_spin_minus}), (\ref{eq:projection_e_flavor}) and (\ref{eq:projection_mu_flavor}), respectively, we obtain:
\begin{align}
\label{eq:direct_left_e}
\mathcal{P}^{(e)}_{L}(t) = 
& c_\nu^2 \left[ r_0^{(1)}(t)- r_3^{(1)}(t) \right] + 
 s_\nu^2\left[ r_0^{(4)}(t)- r_3^{(4)}(t) \right]\nonumber 
\\
&  +c_\nu s_\nu \left[ r_0^{(2)}(t) + r_0^{(3)}(t) -r_3^{(2)}(t) - r_3^{(3)}(t)\right]
\nonumber\\
= & r_0^{(e)}(t) - r_3^{(e)}(t),
\end{align}
where
\begin{align}
\label{eq:re_definitions}
r^{(e)}_i (t) =  c_\nu^2 r^{(1)}_i(t) +s_\nu^2 r^{(4)}_i(t)  + 
c_\nu s_\nu\left[ r^{(2)}_i(t)  +  r^{(3)}_i(t) \right],\nonumber\\
\end{align}
for $i=0,1,2,3$. For the case of measuring active muon neutrinos we get
\begin{equation}
\label{eq:direct_left_mu}
\mathcal{P}^{(\mu)}_{L}(t) =r_0^{(\mu)}(t) - r_3^{(\mu)}(t),
\end{equation}
where
%
\begin{align}
\label{eq:rmu_definitions}
r^{(\mu)}_i (t) =  s_\nu^2 r^{(1)}_i(t) + c_\nu^2 r^{(4)}_i(t) -
c_\nu s_\nu\left[ r^{(2)}_i(t)  +  r^{(3)}_i(t) \right].
\nonumber\\
\end{align}

%

The result in the case of the first measurement scheme~(\ref{direct spin measurement4}) has the form:
\begin{align}
\mathcal{Q}^{(\ell)}_{L}
=\mr{Tr}\{\hat{\Pi}_{\nu_\ell^L}\mathcal
{\hat{F}}[\hat{\varrho}_{\mr{post}}]\} 
=\mr{Tr}\{  \hat \Pi_-  \hat{\Pi}_{\ell}\hat{\varrho}_\mr{post}(t)
\hat{\Pi}_{\ell}\hat \Pi_-\}  .
\end{align}
The final expressions are formally identical to Eqs.~(\ref{eq:direct_left_e}) and~(\ref{eq:direct_left_mu}). However, the functions related to the quadrants of the density matrix, $\rho^{(\alpha)}_i(t)$, must be replaced by the functions corresponding to the post-measurement matrix $\hat \varrho_{\mr{post}}(t)$.

Since at $t=0$ only the right-handed electron neutrino component is different from zero, the functions $r_{\mr{post}, 0}^{(1)}(t)$, $r_{\mr{post}, 3}^{(1)}(t)$ obey the same equation of motion, but with different initial conditions:
\begin{equation}
\label{eq:rpost_t0}
r_{\mr{post}, 0}^{(1)}(0) = r_{\mr{post}, 3}^{(1)}(0) = 
\frac{ r_0^{(1)}(0) + r_3^{(1)}(0)}
{ 1 + 2 (r_3^{(1)}(0)+r_3^{(4)}(0))}.
\end{equation}
The same holds for $r_{\mr{post}, 0}^{(4)}(0)$ and $r_{\mr{post}, 3}^{(4)}(0)$.
Note that the sum in the parenthesis in the denominator can be presented as
\begin{equation}
\label{eq:r3sum}
r_3^{(1)}(0)+r_3^{(4)}(0) = r_3^{(e)}(0)+r_3^{(\mu)}(0).
\end{equation}
Hence
\begin{align}
\label{eq:rpost_electron}
r_{\mr{post},0}^{(e)}(0)=&
c_\nu^2 r_{\mr{post},0}^{(1)}(0)+s_\nu^2 r_{\mr{post},0}^{(4)}(0)
\nonumber\\
&  +s_\nu c_\nu\left[r_{\mr{post},0}^{(2)}(0) + r_{\mr{post},0}^{(3)}(0)\right],
\end{align}
where
\begin{equation}
\label{eq:r23post}
r_{\mr{post},0}^{(2)}(0) + r_{\mr{post},0}^{(3)}(0) =
\frac{r_0^{(2)}(0) + r_0^{(3)}(0)}
{1 + 2 ( r_3^{(e)}(0)+r_3^{(\mu)}(0))}.
\end{equation}

Using the derived results for $\mathcal{P}^{(e)}_{L}$ and $\mathcal{Q}^{(e)}_{L}$ in the expression for the quantum witness~(\ref{quantum witness2}),
%
 we get:
\begin{align}
\label{eq:WL_result}
\mathcal{W}_{L}^{(e)} =
\left|r^{(e)}_0(t) - r^{(e)}_3(t) -
r_{\mr{post}, 0}^{(e)}(t) +r_{\mr{post}, 3}^{(e)}(t)\right|.
\end{align}
%
%

The time evolution of the density matrix $\hat \varrho$ and factors $r^{(\alpha)}_i$ follow the solution of the equation of neutrino motion and Novikov's map in Eq.~(\ref{master equation}). The details of solving analytically the Lindblad equation and, in particular, deriving the minors of the density matrix $\hat\varrho$ in the eigenbasis of the Hamiltonian~(\ref{Hamiltonian of the problem}) are presented in Appendices~\ref{sec:solution_R1} and~\ref{sec:varrho2}. 


\section{Illustration of the invasive effect}
\label{sec:AnalysisAndConclusions}
In the present work, we study the effect of two sequential measurements
done on neutrinos traversing the interstellar space. Due to the cosmic magnetic fields, propagation of a massive
neutrino is accompanied by spin-flavor oscillations, while the stochastic
component of these fields has a random influence on neutrino spin polarization and
leads to the decoherence effect.

We aim at examining the invasive effect of the first, blind measurement, namely the effective neutrino spin filtering by the cosmic object, on the result of the second measurement of an active neutrino performed in the lab on the earth. For this purpose we consider two measurement schemes: we filter out left-handed neutrinos with the cosmic object and then measure its active flavor state at the lab, or measure the active flavor state directly, without spin filtering. Note that the measurement that performs the cosmic object is blind because its result is not recorded. For classical systems, the first
blind measurement is always noninvasive. The difference between results recorded in the lab in the cases of the first and second neutrino propagation schemes is entirely a quantum phenomenon and we quantify it through the quantum witness~(\ref{quantum witness2}).

Below we illustrate the invasive effect of the first, blind measurement, assuming $\mu_{11}=\mu_{22}=\mu_{12}=\mu_\nu$ and that the energy of the neutrino magnetic moment interaction with an interstellar magnetic field is $\mu_\nu B=10^{-32}$~eV. This energy corresponds to the putative magnetic moment, $\mu_\nu\approx4\times10^{-20}\mu_B$, that agrees with the value predicted for the Dirac neutrino by the minimally extended standard model~\cite{Fujikawa:1980prl},
$$
\mu_\nu\approx3.2\times10^{-19}\left(\frac{m_\nu}{1~{\rm eV}}\right)\mu_B,
$$
taking into account current upper bounds on the neutrino mass $m_\nu$ (see, for instance, Refs.~\cite{Cosmology:2019prl,KATRIN:2019prl}). 
When describing the influence of the stochastic magnetic field $\vec{h}$ in the Lindblad master equation~(\ref{master equation}), without loss of generality, we exploit the parametrization (\ref{parametrization}) of the 
dissipator term with a unit vector length $v=1$ and a zero angle $\alpha=0$. Further, from Eq.~(\ref{parametrization}) it follows that if $\alpha=0$ the 
matrix $\hat{V}$ given by Eq.~(\ref{Vmatrix}) is real, and only $v_{1,3}$ components 
enter in the dissipator term, which now depends on the parameter $w^2$ and the angle $\beta$. The Lindblad equation parameter $w^2$ characterizes the strength of the dissipation and is usually equal to some fraction of the energy of the interaction with the magnetic field $\vec{B}$. Here we use a reasonable value $w^2=0.1\mu_\nu B$. For the angle $\beta$ we use a value of $\pi/4$, implying that both neutrino longitudinal and transverse spin components are thermalized due to the stochastic magnetic field on equal footing.  
\begin{figure}[h]
\includegraphics[width=0.45\textwidth]{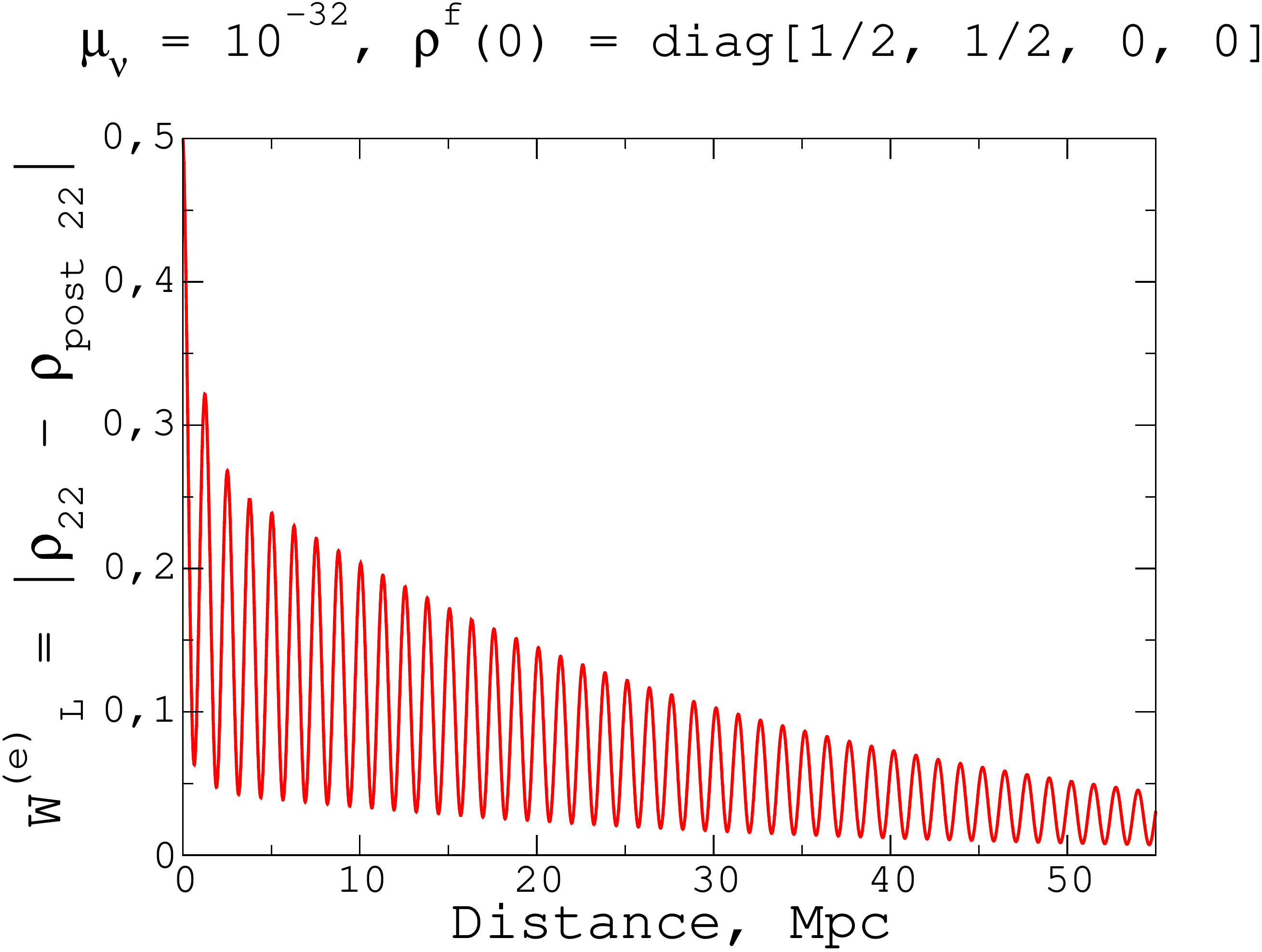}
\caption{The quantum witness for detecting left-handed electron neutrinos at the lab depending on the distance from the cosmic object to the earth. 
In the initial state, there are only electron neutrinos with equal fractions of left- and right-handed particles: $r_0^{(e)}(0) =  1/2$, $r_3^{(e)}(0) =  0$. 
}\label{fig:7}
\end{figure}
\begin{figure}[h]
\includegraphics[width=0.45\textwidth]{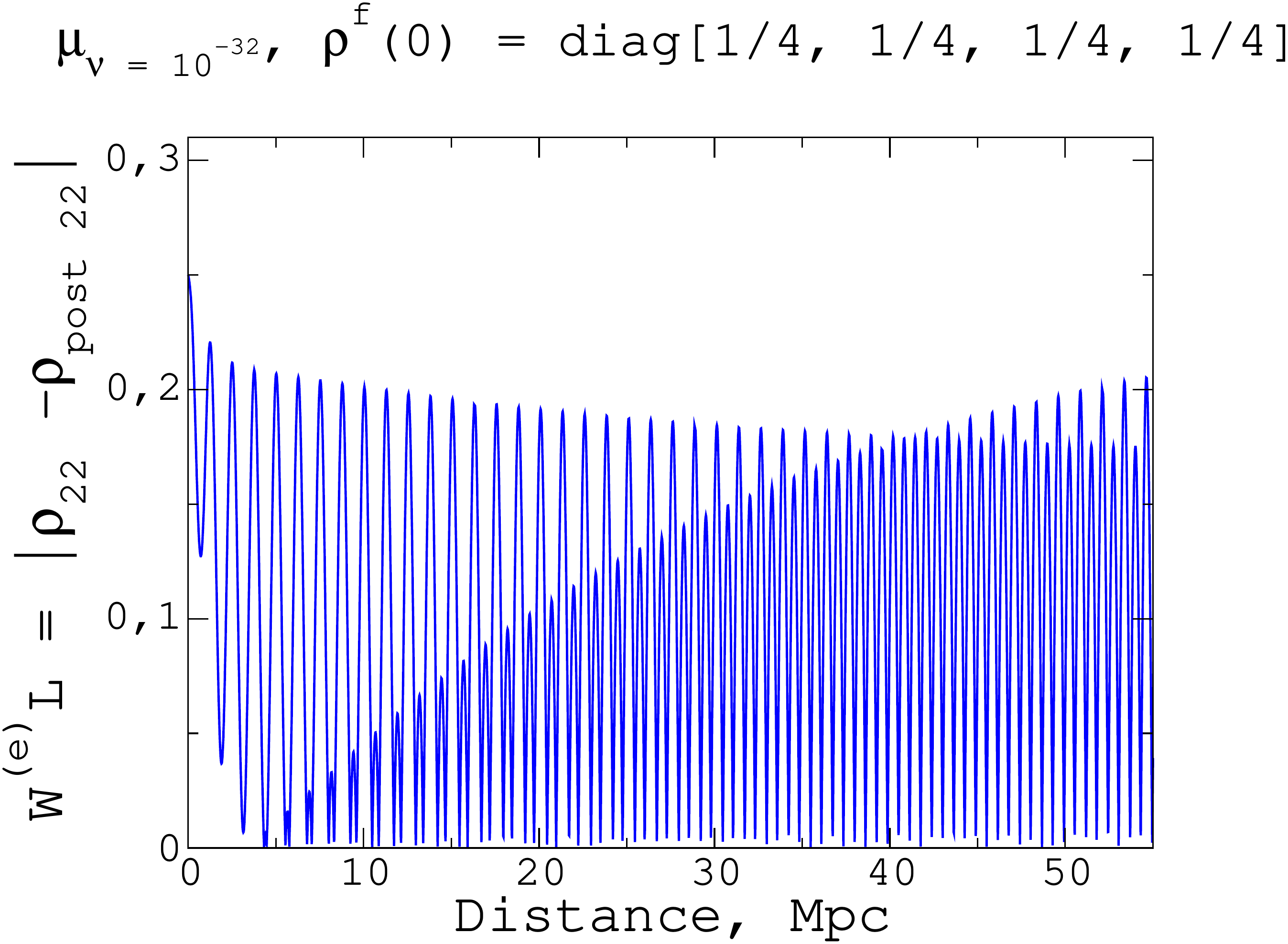}
\caption{The same as in fig.~\ref{fig:7}, but when in the initial state the density matrix in the flavor basis is given by $\hat\varrho={\rm diag}(1/4,1/4,1/4,1/4)$.}\label{fig:8}
\end{figure}
Figure~\ref{fig:7} shows the quantum witness~(\ref{eq:WL_result}) as a function of the distance between the cosmic object and the earth when the neutrino flux at $t=0$ (i.e., at the moment of time corresponding to the neutrino passing through the cosmic object in the first propagation scheme) consists only of electron neutrinos. It is assumed in Fig.~\ref{fig:7} that the neutrinos have already traveled a large distance in the interstellar space before meeting the cosmic object, so that their spins have been fully thermalized. For this reason, we set $r_3^{(e)}(0)=0$, meaning that the numbers of right- and left-handed neutrinos entering the cosmic spin filter are equal. The shown quantum witness function decays, exhibiting oscillations, and tends to zero at large distances. The latter means that the invasive effect of the cosmic object on the result of the measurement performed in the lab vanishes. Such a decaying behavior of the quantum witness is not general. In Fig.~\ref{fig:8} we show the quantum witness~(\ref{eq:WL_result}) when the density matrix describing the neutrino flux before entering the cosmic spin filter is fully thermalized in the flavor basis. It can be seen that the quantum witness function has now a more complex behavior, which is different from ``quantum beats'' observed in Fig.~\ref{fig:7}. First, it decays only at small distances. Second, it also exhibits oscillations, but the oscillation pattern shows a superposition of two functions that oscillate with the same or almost the same frequency but in antiphase: the one function is decaying and the other is growing. The observed picture is explained by the behaviors of the probabilities $\mathcal{Q}^{(e)}_L$ and $\mathcal{P}^{(e)}_L$  of measuring the active electron neutrino at the lab in the first and second propagation schemes, respectively.

\section{Summary}
\label{summary}
In this work we have developed a new approach for interpretation and analysis of the data of experiments with cosmic neutrinos. Our approach employs the concepts of invasiveness and quantum witness that are used in the theory of quantum measurements. We have considered two subsequent measurements of neutrinos from a distant source traversing the interstellar space and interacting with stochastic interstellar magnetic fields. The first measurement is performed by a dense cosmic object, which effectively absorbs active, left-handed-neutrinos, thus letting only right-handed neutrinos to propagate further to the earth. The second measurement is performed in the lab at the earth and is a typical measurement of active flavor neutrinos. The first measurement is blind and has an invasive effect on the result of the second measurement, undermining the credibility of our physical interpretation of the results obtained in the lab. In order to quantify this credibility, we defined the quantum witness for the neutrino measurement in the lab and obtained analytical expressions for the quantum witness in terms of the elements of the neutrino density matrix. Using an anlytical solution of the Lindblad master equation for the neutrino density matrix, we demonstrated the invasive effect of the cosmic filter on the neutrino measurement in the lab. We showed that the quantum witness can exhibit qualitatively different behaviors for different properties of the neutrino flux that meets the cosmic object. In particular, the decay of the quantum witness as a function of the distance from the cosmic object that one may naively expect due to the neutrino spin thermalization effects is not observed when the density matrix of the neutrino flux meeting the cosmic object has in the flavor basis a diagonal form with all four neutrino states equally populated. Our approach can be used for quantifying theoretical uncertainties associated with the lack of information about the conditions of cosmic neutrinos propagation when interpreting and analyzing the results of experiments with neutrinos from distant astrophysical sources.

\begin{acknowledgments}
The work of K.A.K. and A.I.S. is supported by the Russian Foundation for Basic Research under grant no. 20-52-53022-GFEN-A.
\end{acknowledgments}

\appendix

\section{The Lindblad master equation}
\label{sec:app_lindblad}
In what follows, we use the eigenbasis $(\tilde{\nu}_1,\tilde{\nu}_2,\tilde{\nu}_3,\tilde{\nu}_4)^T$ of the Hamiltonian~(\ref{Hamiltonian of the problem}) (see Ref.~\cite{Kurashvili}). In this representation the Lindblad master equation~(\ref{master equation}) takes the form 
\begin{align}
\label{eq:lindblad}
\frac{d\rho_{nm}}{dt} =& -i (E_n - E_m)\varrho_{mn} - \frac{w^2}{2}
\sum_q \left(\rho_{nq} V^2_{qm} + V^2_{nq} \varrho_{qm}\right)
\nonumber \\
&+ w^2 \sum_{q,s} V_{nq} \varrho_{qs} V_{sm},
\end{align}
where $E_{1,2,3,4}$ are the eigenvalues of the Hamiltonian~(\ref{Hamiltonian of the problem}).

We present the density matrix as
\begin{equation}
\label{eq:dmatrix_R}
\hat\varrho =
\begin{pmatrix}
\varrho^{(1)} & \varrho^{(2)} \\
\varrho^{(3)} & \varrho^{(4)}
\end{pmatrix},
\end{equation}
where $\varrho^{(\alpha)}$ are the $2 \times 2 $ minors.

Then in Eq.~(\ref{eq:lindblad}) one has the expressions
of the form
\begin{equation}
\label{eq:Vsqr}
\sum_{q=1}^2  ({v}_{nq}^{(\nu_\ell)2} \varrho_{qm}^{(\alpha)} + R^{(\alpha)}_{nq}v_{qm}^{(\nu_\ell)2}),
\end{equation}
and
\begin{equation}
\label{eq:VAV}
\sum_{q,s=1}^2 v_{nq}^{(\nu_\ell)} R^{(\alpha)}_{qs} v_{sm}^{(\nu_\ell)}
\end{equation}
for the first and second sums, respectively. All $\varrho^{(\alpha)}$ along with
$v^{(\nu_\ell)}$ are expanded in the basis of the three Pauli matrices and $2 \times 2$
unit matrix:
\begin{equation}
\label{eq:Rsigma}
R^{(\alpha)} = r^{(\alpha)}_0 I + \vec r^{(\alpha)} \cdot \vec \sigma,
\end{equation}
and
\begin{equation}
\label{eq:vsigma}
v^{(\nu_\ell)} = v_0 I + \vec v \cdot \vec \sigma.
\end{equation}
with
\begin{align}
\label{eq:R_expand}
r^{(\alpha)}_0 = \frac{1}{2}(\varrho^{(\alpha)}_{11} + \varrho^{(\alpha)}_{22}),\,\\
r^{(\alpha)}_1 = \frac{1}{2}(\varrho^{(\alpha)}_{12} + \varrho^{(\alpha)}_{21}),\,\\
r^{(\alpha)}_2 = \frac{i}{2}(\varrho^{(\alpha)}_{12} - \varrho^{(\alpha)}_{21}),\,\\
r^{(\alpha)}_3 = \frac{1}{2}(\varrho^{(\alpha)}_{11} - \varrho^{(\alpha)}_{22}).
\end{align}
Analogous relations are used between the elements of the matrix $v$ and
coefficients of its expansion.

We now expand Eqs.~(\ref{eq:Vsqr}) and~(\ref{eq:VAV}) and exploit
Eqs.~(\ref{eq:vsigma}) and~(\ref{eq:Rsigma}):
\begin{align}
\label{eq:Vsqr_expand}
\sum_{q=1}^2 & \left(
v_{nq}^{(\nu_\ell)2} \varrho_{qm}^{(\alpha)} + \varrho^{(\alpha)}_{nq}v_{qm}^{(\nu_\ell)2} \right) 
 \nonumber\\
=  &     2 \left[ \left( v_0^2 + v^2 \right)r^{(\alpha)}_0 +
2 v_0 \vec v \cdot \vec r^{(\alpha)} \right] I_{nm} 
 \nonumber\\
&+ \, 4 v_0 r^{(\alpha)}_0 \vec v \cdot \vec \sigma_{nm} +
2 \left( v_0^2 + v^2 \right) \vec r^{(\alpha)} \cdot \vec \sigma_{nm},
\end{align}
\begin{align}
\label{eq:VAV_expand}
\sum_{q,s=1}^2 & v_{nq}^{(\nu_\ell)} \varrho^{(\alpha)}_{qs} v_{sm}^{(\nu_\ell)} 
\nonumber\\
 = &\left[ \left( v_0^2 + v^2 \right)r^{(\alpha)}_0
+ 2 v_0 \vec v \cdot \vec r^{(\alpha)} \right] I_{nm} 
 \nonumber\\
& +2 \left[ v_0 r^{(\alpha)}_0 + \vec v \cdot \vec r^{(\alpha)}\right]
\vec v \cdot \vec \sigma_{nm}
 \nonumber\\
& + \left( v_0^2 - v^2 \right) \vec r^{(\alpha)}\cdot \vec \sigma_{nm}.
\end{align}

Summing up Eqs.~(\ref{eq:Vsqr_expand}) and~(\ref{eq:VAV_expand})
with the same weights as in Eq.~(\ref{eq:lindblad}), one gets the following dissipative contribution:
\begin{align}
\label{eq:dissipative}
L^{(\alpha)} =&
-\frac{w^2}{2}
\sum_{q=1}^2 \left(
v_{nq}^{(\nu_\ell)2} \varrho_{qm}^{(\alpha)} + \varrho^{(\alpha)}_{nq}v_{qm}^{(\nu_\ell)2}
\right) 
 \nonumber\\
&+ w^2 \sum_{q,s=1}^2 v_{nq}^{(\nu_\ell)} \varrho^{(\alpha)}_{qs} v_{sm}^{(\nu_\ell)}
\nonumber\\
=&  2 w^2 \left[ (\vec v \cdot \vec r^{(\alpha)})
\vec v \cdot \vec \sigma_{nm} -
v^2  \vec r^{(\alpha)} \cdot \vec \sigma_{nm}\right].
\end{align}
Eq.~(\ref{eq:dissipative}) is also decomposed in the
basis of $2 \times 2$ matrices:
\begin{equation}
      \label{eq:Ldecompose}
      L^{(\alpha)} =  \Lambda^{(\alpha)}_0 I +
      \vec \Lambda^{(\alpha)} \cdot \vec \sigma,
\end{equation}
where
\begin{align}
\label{eq:l0}
\Lambda^{(\alpha)}_0 & = 0, \\
\label{eq:l1}
\Lambda^{(\alpha)}_i & = 2 w^2
\left[(\vec v \cdot \vec r^{(\alpha)}) v_i - v^2 r_i^{(\alpha)} \right].
\end{align}

Let us derive the equations for the elements of the minor
$\varrho^{(1)}$. Using Eqs.~(\ref{eq:Rsigma}), (\ref{eq:R_expand}), (\ref{eq:Vsqr_expand}), (\ref{eq:VAV_expand}), (\ref{eq:dissipative}), and~(\ref{eq:lindblad}), one gets
\begin{align}
\label{eq:diff_rho11}
\frac{d}{dt}\varrho_{11} & =
\frac d {dt} (r^{(1)}_0 + r^{(1)}_3) = \Lambda^{(1)}_3,
\\
\label{eq:diff_rho22}
\frac{d}{dt} \varrho_{22}& =
\frac d {dt} (r^{(1)}_0 - r^{(1)}_3)= -\Lambda^{(1)}_3,
\\
\label{eq:diff_rho12}
\frac d {dt} \varrho_{12} & =
\frac d {dt} r^{(1)}_- = - i\omega_{12} r^{(1)}_- + \Lambda^{(1)}_-,
\\
\label{eq:diff_rho21}
\frac d {dt} \varrho_{21} & =
\frac d {dt} r^{(1)}_+ = - i \omega_{21} r^{(1)}_+ + \Lambda^{(1)}_+,
\end{align}
where $r_\pm = r_1 \pm i r_2$, $\Lambda^{(1)}_\pm=\Lambda^{(1)}_1\pm i\Lambda^{(1)}_2$,
and $\omega_{12} = E_1 - E_2 = -\omega_{21}$. Note that the sum of the diagonal matrix elements
$\varrho_{11} + \varrho_{22} = r^{(1)}_0$ is constant in time.
The set of equations for $\varrho^{4}$ is obtained from
Eqs.~(\ref{eq:diff_rho11})-(\ref{eq:diff_rho21}) 
by changing $r^{(1)}$ to $r^{(4)}$ and $\omega_{12}$ to $\omega_{34}$.
The sum $\varrho_{33} + \varrho_{44} = r^{(4)}_0$
is also conserved as well as the complete
trace of the density matrix.

The set of equations for the minor $\varrho^{(2)}$ reads
\begin{align}
\label{eq:diff_rho13}
\frac d {dt} \varrho_{13}  =&
\frac d {dt} (r^{(2)}_0 + r^{(2)}_3)
\nonumber\\
=& - i \omega_{13}(r^{(2)}_0 + r^{(2)}_3)
+ \Lambda^{(2)}_3,
\\
\label{eq:diff_rho24}
\frac d {dt} \varrho_{24} =&
\frac d {dt} (r^{(2)}_0 - r^{(2)}_3) 
 \nonumber\\
=& - i \omega_{24}(r^{(2)}_0 - r^{(2)}_3)  - \Lambda^{(2)}_3,
\\
\label{eq:diff_rho14}
\frac d {dt} \varrho_{14} =&
\frac d {dt} r^{(2)}_- =
- i \omega_{14} r^{(2)}_-  + \Lambda^{(2)}_-
\\
\label{eq:diff_rho23}
\frac d {dt} \varrho_{23} & =
\frac d {dt} r^{(2)}_+ =
- i\omega_{23} r^{(2)}_+  + \Lambda^{(2)}_+.
\end{align}
The equations for $\varrho^{(3)}$ are obtained from Eqs.~(\ref{eq:diff_rho13})-(\ref{eq:diff_rho23}) upon Hermitian
conjugation.

\section{The solution for  $\varrho^{(1)}$}
\label{sec:solution_R1}
Below we consider the system of equations for the minor  $\varrho^{(1)}$, omitting upper indexes in $r^{(\alpha)}$
and designating $\omega = \omega_{12} = - \omega_{21}$. After redefinition of the time variable
\begin{equation}
\label{eq:timeredefine}
\tau =  2 w^2  t,
\end{equation}
the equations acquire the following form:
\begin{align}
\label{eq:diff_r0}
\frac d {d\tau} r_0 = & 0, \\
\label{eq:diff_rplus}
\frac d {d\tau} r_+  =&
[\frac{v_+ v_-} 2 - v_3^2 - i \homega ] r_+
+ \frac{v_+^2} 2 r_- 
\nonumber\\
& + v_+ v_3 r_3, \\
\label{eq:diff_rminus}
\frac d {d\tau} r_- =& 
\frac{v_-^2} 2 r_+
+ [ - \frac{v_+ v_-} 2 -  v_3^2 + i \homega ] r_-
\nonumber\\
& +  v_- v_3 r_3, \\
\label{eq:diff_r3}
\frac d {d\tau} r_3  =&
v_-  v_3 \frac {r_+} 2 +    v_+  v_3 \frac {r_-} 2 - v_+ v_- r_3,
\end{align}
where $\homega = \omega / 2w^2$ .

In the case of $v^2 = 1$ and $v_y = 0$, one has $\vec v = (\sin \beta, 0, \cos \beta)$
and it is convenient to rewrite the system of equations in terms of $r_{1,2}$ instead of $r_\pm$:
\begin{align}
\label{eqn:diff_r0_alt}
\frac d {d\tau} r_0 & = 0,
\\
\label{eqn:diff_r1_bar}
\frac d {d\tau} r_1 & = -r_1\cos^2 \beta  + r_2 \bar \omega  +
r_3 \sin \beta \cos \beta ,
\\
\label{eqn:diff_r2_alt}
\frac d {d\tau} r_2 & = - r_1 \bar \omega - r_2 \cos^2 \beta ,
\\
\label{eqn:diff_r3_alt}
\frac d {d\tau} r_3 & =  r_1 \sin \beta \cos \beta - r_3 \sin^2 \beta.
\end{align}

To solve the above system, one must diagonalize the
$3\times3$ matrix:
\begin{equation}
\label{eq:MR1}
\mathcal{M}^{(1)} =
\begin{pmatrix}
- \cos^2\beta && \homega && \sin\beta \cos\beta
\\
-\homega && - \cos^2\beta  && 0
\\
\sin\beta \cos\beta  && 0 && -\sin^2\beta
\end{pmatrix}.
\end{equation}
%
 The general solution of the system is a sum of exponents:
\begin{equation}
\label{eq:solution_R1}
r_i (\tau) = \sum_{k=1}^{3} C_{ik} e^{i\nu_k \tau},
\end{equation}
where $\nu_i$ are the eigenvalues of the above matrix and
the integration constants are given by the following
expressions:
\begin{align}
\label{eq:R1_c1}
C_{i1} & = \frac {B_{0i} \nu_2 \nu_3 - B_{1i} (\nu_2 + \nu_3)
+ B_{2i}}
{(\nu_1 - \nu_2)(\nu_1 - \nu_3)},\\
\label{eq:R1_c2}
C_{i2} & = \frac{B_{0i} \nu_1 \nu_3 - B_{1i} (\nu_1 +\nu_3)
+B_{2i}}
{(\nu_2 - \nu_1)(\nu_2 - \nu_3)},\\
\label{eq:R1_c3}
C_{i3} & = \frac{B_{0i} \nu_1 \nu_2 - B_{1i} (\nu_1 + \nu_2)
+B_{2i}}
{(\nu_3 - \nu_1)(\nu_3 -\nu_2)},
\end{align}
where
\begin{align}
\label{eq:R1betas}
B_{0i} & = r_i(0), \, \,
\\
B_{1i} & = \sum_{k=1}^3 \mathcal{M}^{(1)}_{ik}r_{k}(0), \, \,
\\
B_{2i} & = \sum_{k,l=1}^3 \mathcal{M}^{(1)}_{ik}
\mathcal{M}^{(1)}_{kl}
r_{l}(0).
\end{align}
The equations for $\varrho^{(4)}$ are similar to Eqs.~(\ref{eq:diff_r0})-(\ref{eq:diff_r3}),
except that one must replace $\omega=\omega_{12}$ with
$\omega=\omega_{34}=E_3-E_4$. 
Note that there is a condition of the trace conservation:
\begin{equation}
\label{eq:trace_unitiy}
\mathrm{Tr}\varrho = \sum_i \varrho_{ii} =
r^{(1)}_0+r^{(4)}_0 = 1.
\end{equation}
Another limitation is
that the density matrix should be Hermitian,
$\varrho_{12} = \varrho^\dag_{21}$, $\varrho_{34} = \varrho^\dag_{43}$,
which is already satisfied
by (\ref{eq:R1_c1})-(\ref{eq:R1betas}).

\section{The solution for $\varrho^{(2)}$}
\label{sec:varrho2}
The system of equations for the minor $\varrho^{(2)}$ is given by
\begin{align}
\label{eq:diff_r0_2}
\frac d {d\tau} r_ 0 =& 
- \frac i 2 (\homega_{13} + \homega_{24}) r_ 0
- \frac i 2 (\homega_{13} - \homega_{24}) r_3,
\\
\label{eq:diff_rp_2}
\frac d {d\tau} r_+ =& 
\left[ -\frac{v_+ v_-} 2 - v_3^2 -i \homega_{23}\right] r_+ 
\frac{v_+^2} 2 r_- +
 \nonumber\\
& + v_+v_3 r_3,
\\
\label{eq:diff_rm_2}
\frac d {d\tau} r_- = &
\frac{v_-^2} 2 r_+ +
\left[ -\frac{v_+ v_-} 2 - v_3^2 - i \homega_{14}\right]r_- 
 \nonumber\\
& + v_- v_3 r_3,
\\
\label{eq:diff_r3_2}
\frac d {d\tau} r_3  =&
- \frac i 2 (\homega_{13} - \homega_{24}) r_ 0
+ \frac{v_- v_3} 2 r_+ 
 \nonumber\\
& +\frac{v_+ v_3} 2 r_- +
\left[ - \frac i 2 (\homega_{13} + \homega_{24})
-v_+ v_-\right]r_3.
\end{align}

Using the ansatz $\vec v  = (\sin \beta , 0, \cos \beta)$
and changing to $r_1$ and $r_2$, one gets the system
for $\varrho^{(2)}$ as
\begin{align}
\label{eqn:diff_r0_2}
\frac d {d\tau} r_ 0  = &- i  \homega_+ r_ 0 - i \homega_- r_3,
\\
\label{eqn:diff_rp_2}
\frac d {d\tau} r_1  =&
( - \cos^2 \beta - i \homega_+) r_1  +
\homega_0 r_2  
 \nonumber\\
& + r_3 \sin \beta \cos \beta,
\\
\label{eqn:diff_rm_2}
\frac d {d\tau} r_2 = &
- \homega_0  r_1 + ( -\cos^2 \beta - i \homega_+)r_2,
\\
\label{eqn:diff_r3_2}
\frac d {d\tau} r_3 = &
-i \homega_- r_0 + r_1 \sin \beta \cos \beta 
 \nonumber\\
& - ( i\omega_+  + \sin^2 \beta)r_3.
\end{align}
The frequencies in the above formulas are
\begin{align}
\label{eqn:freq0}
\homega_0 & = \frac{\homega_{12} + \homega_{34} }{2},
\\
\label{eqn:freqpm}
\homega_\pm & = \frac{\homega_{13} \pm \homega_{24} }{2}.
\end{align}

Solving the system requires diagonalization of the
$4\times 4$ matrix:
\begin{widetext}
\begin{equation}
\label{eqn:MR2}
\mathcal{M}^{(2)}_4 =
\begin{pmatrix}
-i \homega_+ && 0 && 0 && -i \homega_-
\\
0 && -\cos^2\beta - i \homega_+ && \homega_0 && \sin\beta\cos\beta
\\
0 && -\homega_0 && -\cos^2\beta- i \homega_+ && 0
\\
-i \homega_- && \sin\beta\cos\beta && 0 && -\sin^2\beta - i \omega_+
\end{pmatrix}.
\end{equation}
The final solution is given in form of a linear combination
of exponential functions:
\begin{equation}
\label{eqn:solutionR1R2}
r^{(2)}_i (\tau) = \sum_{k=1}^4 C_{ik}e^{\nu_k \tau},
\end{equation}
with the integration constants
\begin{align}
\label{eqn:R2_Ci1}
C_{i1} & = \frac{- B_{0i}\nu_2 \nu_3 \nu_4
+ B_{1i}(\nu_2\nu_3 + \nu_2\nu_4 +\nu_3\nu_4)
- B_{2i}(\nu_2 + \nu_3 + \nu_4) + B_{3i}}
{(\nu_1 -\nu_2)(\nu_1-\nu_3)(\nu_1-\nu_4)}, \\
\label{eqn:R2_Ci2}
C_{i2} & = \frac{ B_{0i}\nu_1 \nu_3 \nu_4
- B_{1i}(\nu_1\nu_3 + \nu_1\nu_4 +\nu_3\nu_4)
+ B_{2i}(\nu_1 + \nu_3 + \nu_4) + B_{3i}}
{(\nu_2 -\nu_1)(\nu_2-\nu_3)(\nu_2-\nu_4)}, \\
\label{eqn:R2_Ci3}
C_{i3} & = \frac{- B_{0i}\nu_1 \nu_2 \nu_4
+ B_{1i}(\nu_1\nu_2 + \nu_1\nu_4 +\nu_2\nu_4)
- B_{2i}(\nu_1 + \nu_2 + \nu_4) + B_{3i}}
{(\nu_3 -\nu_1)(\nu_3-\nu_2)(\nu_3-\nu_4)}, \\
\label{eqn:R2_Ci4}
C_{i4} & = \frac{ B_{0i}\nu_1 \nu_2 \nu_3
- B_{1i}(\nu_1\nu_2 + \nu_1\nu_3 +\nu_2\nu_3)
+ B_{2i}(\nu_1 + \nu_2 + \nu_3) + B_{3i}}
{(\nu_4 -\nu_1)(\nu_4-\nu_2)(\nu_4-\nu_1)},
\end{align}
where
\begin{align}
\label{eqn:betas_R2}
B_{0i} &= r_i(0),
\\
B_{1i} &= \sum_{k=1}^3 \mathcal{M}^{(2)}_{ik} r_k(0),
\\
B_{2i} &= \sum_{k,l=1}^3 \mathcal{M}^{(2)}_{ik} \mathcal{M}^{(2)}_{kl} r_l(0),
\\
B_{3i} &= \sum_{k,l,m=1}^3 \mathcal{M}^{(2)}_{ik} \mathcal{M}^{(2)}_{kl}
\mathcal{M}^{(2)}_{lm} r_m(0).
\end{align}

\end{widetext}

\begin{thebibliography}{99}
%
\bibitem{Zhang2015}
    H. Zhang et al. [The Super-Kamiokande Collaboration], Astropart. Phys.
    \textbf{60}, 41 (2015).
%
\bibitem{Hirata1987}
    K. Hirata et al. [Kamiokande-II Collaboration], Phys. Rev. Lett. \textbf{58}, 1490 (1987); R. M. Bionta et al., Phys. Rev. Lett. \textbf{58}, 1494 (1987).
%
\bibitem{Kaspi2017}
    V. M. Kaspi, A. M. Beloborodov, Ann. Rev. Astron. Astrophys. \textbf{55} 261 (2017).
\bibitem{Abbott2020}
    B. P. Abbott et al. [LIGO Scientific Collaboration], [Virgo Collaboration],
    Astrophys. J. Letters \textbf{892}, L3 (2020).
\bibitem{Wang2007} X.-Y. Wang, S. Razzaque, P. Meszaros, Z.-G. Dai,
    Phys. Rev. D \textbf{76}, 083009 (2007).
\bibitem{Aab2020}
    A. Aab, et al. JCAP (submitted) arXiv:2004.10591
\bibitem{Kalashev2015}
    O. Kalashev, D. Semikoz, I. Tkachev,
    JETP \textbf{120}, 541 (2015).
\bibitem{Kollatschny2015}
    W. Kollatschny, N. Schartel, M. Zetzl, M. Santos-Lleo, P. M. Rodriguez-Pascual, and L.
    Ballo, Astron. Astrophys. \textbf{577}, L1 (2015).
\bibitem{Sobey2019}
    C. Sobey et al. Mon. Not. Roy. Astron. Soc. \textbf{484}, 3646 (2019).
\bibitem{Ade2016}
    P. A. R. Ade et al. [Planck Collaboration],
    Astron. Astrophys. \textbf{594}, A19 (2016).
\bibitem{Brandenburg2017}
    A. Brandenburg et al. Phys. Rev. D, \textbf{96} 123528 (2017).
%
\bibitem{bib:Giunti} C. Giunti and A. Studenkin, Rev. Mod. Phys. \textbf{87},
531 (2015).
%
\bibitem{andp16} C. Giunti, K. A. Kouzakov, Y.-F. Li, A. V. Lokhov, A. I. Studenikin, and S. Zhou, {Ann. Phys. (Berlin)} \textbf{528}, 198 (2016).
%
\bibitem{Deffner}
S. Deffner, Phys. Rev. E \textbf{88}, 062128 (2013).
%
\bibitem{Formaggio}
J. A. Formaggio, D. I. Kaiser, M. M. Murskyj, T. E. Weiss, Phys. Rev. Lett. \textbf{117}, 050402 (2016).
%
\bibitem{Shibata}
M. Ban, S. Kitajima, and F. Shibata, Phys. Rev. A \textbf{97}, 052101 (2018).
%
\bibitem{Novikov} E. A. Novikov, JETP \textbf{20}, 1290 (1965); A. Dutta, A.
    Rahmani, and A. del Campo, Phys. Rev. Lett. \textbf{117}, 080402 (2016).
%
\bibitem{EmaryLambertNori}
C. Emary, N. Lambert and F. Nori,  Rep. Prog. Phys. \textbf{77}, 039501 (2014).
\bibitem{HoffmannSpee} J. Hoffmann, C. Spee, O. G\"uhne, and C. Budroni,   New J. Phys. \textbf{20}, 102001 (2018).
\bibitem{KoflerBrukner} J. Kofler and C. Brukner, Phys. Rev. A \textbf{87}, 052115 (2013).
\bibitem{HuangJiajieLing}
Xue-Ke Song, Yanqi Huang, Jiajie Ling, and Man-Hong Yung, Phys. Rev. A \textbf{98}, 050302(R) (2018).
\bibitem{HalliwellMawby} J. J. Halliwell and C. Mawby,  Phys. Rev. A \textbf{100}, 042103 (2019).
\bibitem{BanerjeeJayannavar}
J. Naikoo, S. Banerjee, and A. M. Jayannavar,  Phy. Rev. A \textbf{100}, 062132 (2019).
\bibitem{NikitinToms}
N. Nikitin and K. Toms,  Phys. Rev. A \textbf{100}, 062314 (2019).
\bibitem{NathanWilliamsAndrew}
N. S. Williams and Andrew N. Jordan, Phys. Rev. Lett. \textbf{100}, 026804 (2008).
\bibitem{MeschedeEmary}
C. Robens, W. Alt, D. Meschede, C. Emary, and A. Alberti, Phys. Rev. X \textbf{5}, 011003 (2015).
\bibitem{SkinnerRuhmanNahum}
B. Skinner, J. Ruhman, and A. Nahum,  Phys. Rev. X \textbf{9}, 031009 (2019).
\bibitem{Leggett} A. J. Leggett and A. Garg, Phys. Rev. Lett. \textbf{54}, 857 (1985);
G. Schild and C. Emary Phys. Rev. A \textbf{92}, 032101 (2015); K. Wang, G. C. Knee, X. Zhan, Z. Bian, J. Li, and P. Xue
Phys. Rev. A \textbf{95}, 032122 (2017); D. Avis, P. Hayden, and M. M. Wilde Phys. Rev. A \textbf{82},
030102(R) (2010). C. M. Li, N. Lambert, Y. N. Chen, G. Y. Chen, and F.
Nori, Sci. Rep. \textbf{2}, 885 (2012).
%
%
\bibitem{Lindblad} G. Lindblad, Comm. Math. Phys. \textbf{48(2)},119 (1976).
\bibitem{Stankevich:2020icp}
K.~Stankevich and A.~Studenikin,
Phys. Rev. D \textbf{101} 056004 (2020).
\bibitem{Farzan_Schwetz_Smirnov} Y. Farzan, T. Schwetz, A. Yu. Smirnov, JHEP \textbf{0807} (2008) 067.
\bibitem{Lisi_Marrone} E. Lisi, A. Marrone, D. Montanino, Phys. Rev. Lett. \textbf{85}, 1166 (2000). 
\bibitem{Barenboim_Mavromato} G. Barenboim, N. E. Mavromato, JHEP \textbf{0501}, 034 (2005).
\bibitem{Barenboim_Mavromatos2} G. Barenboim, N. Mavromatos, S. Sarkar,  A. Waldron-Lauda, Nucl. Phys. B \textbf{758}, 90 (2006).
\bibitem{Benatti_Floreanini} F. Benatti, R. Floreanini, Phys. Rev. D \textbf{64}, 085015 (2001).
\bibitem{Oliveira2014} R. L. N. Oliveira, M. M. Guzzo, P. de Holanda, Phys. Rev. D \textbf{89},  053002 (2014).
\bibitem{Oliveira2016}	R. L. N. Oliveira, Eur. Phys. J. C \textbf{76}, 417 (2016).	
\bibitem{Balieiro_Guzzo} G. Balieiro Gomes, M.M. Guzzo, P.C. de Holanda, R. L. N. Oliveira, Phys. Rev. D \textbf{95}, 113005 (2017).
\bibitem{Joao_Coelho} Joao A.B. Coelho, W. Anthony Mann, Phys. Rev. D \textbf{96}, 093009 (2017).
\bibitem{Joao_Coelho2} Joao A. B. Coelho, W. Anthony Mann, Saqib S. Bashar, Phys. Rev. Lett. \textbf{118}, 221801 (2017).
\bibitem{Capolupo_Giampalo} A. Capolupo, S.M. Giampaolo, G. Lambiase. e-Print: arXiv:1807.07823 [hep-ph] (2018).
\bibitem{Coloma} P. Coloma, J. Lopez-Pavon, I. Martinez-Soler, H. Nunokawa, Eur. Phys. J. C \textbf{78}, 614 (2018).
%
\bibitem{Kurashvili} P. Kurashvili, K. A. Kouzakov, L. Chotorlishvili, and A.
    I. Studenikin Phys. Rev. D \textbf{96}, 103017 (2017).
%
\bibitem{bib:Fabbricatore} R. Fabbricatore, A. Grigoriev, and A. Studenikin, J. Phys. Conf. Ser. \textbf{718}, 062058 (2016).
%
\bibitem{Garanin:1997prb} D. A. Garanin, Phys. Rev. B. \textbf{55}, 3050 (1996).
%
\bibitem{Fujikawa:1980prl} K. Fujikawa,  R. Shrock, Phys. Rev. Lett. \textbf{45}, 963 (1980).
%
\bibitem{Cosmology:2019prl} A. Loureiro \emph{et al.}, Phys. Rev. Lett. \textbf{123}, 081301 (2019).
%
\bibitem{KATRIN:2019prl} M. Aker \emph{et al.} (KATRIN Collaboration), Phys. Rev. Lett. \textbf{123}, 221802 (2019).
%
\bibitem{PopovStudenikin}
A. Popov, A. Studenikin, Eur. Phys. J. C \textbf{79}, 144 (2019).
%
%
\end{thebibliography}
\end{document}